\documentclass{sig-alternate-05-2015}
\usepackage{paralist}
\usepackage{times}
\usepackage{latexsym}
\usepackage{amsmath,amssymb}
\usepackage{url}
\usepackage{color}
\usepackage{graphicx}
\usepackage{algorithm}
\usepackage{algpseudocode}
\usepackage{epstopdf}
\usepackage{dsfont,pifont}
\usepackage{bbm}
\usepackage{booktabs}
\usepackage[english]{babel}
\usepackage{framed}
\usepackage{multicol,multirow}
\usepackage{CJK}
\usepackage{indentfirst}
\usepackage{amsmath}
\usepackage{caption}
\usepackage{subcaption}

\usepackage[bookmarks=true%
,bookmarksnumbered=true%
,hypertexnames=false%
,breaklinks=true%
,colorlinks=true%
,linkcolor=blue%
,citecolor=blue%
,urlcolor=blue%
]{hyperref}
\usepackage{hyperref}

\def\ignore#1{}

\begin{document}


\CopyrightYear{2016} 
\setcopyright{acmcopyright}
\conferenceinfo{CIKM'16 ,}{October 24-28, 2016, Indianapolis, IN, USA}
\isbn{978-1-4503-4073-1/16/10}\acmPrice{\$15.00}
\doi{http://dx.doi.org/10.1145/2983323.2983818}

\clubpenalty=10000 
\widowpenalty = 10000

%
%
%
%
%
%

\title{aNMM: Ranking Short Answer Texts with Attention-Based Neural Matching Model} 
\author{
\alignauthor
Liu Yang$^1$ \quad  Qingyao Ai$^1$ \quad Jiafeng Guo$^2$ \quad  W. Bruce Croft$^1$\\
       \affaddr{$^1$ Center for Intelligent Information Retrieval, University of Massachusetts Amherst, MA, USA \\
	$^2$ CAS Key Lab of Network Data Science and Technology, Institute of Computing Technology,\\
Chinese Academy of Sciences, Beijing, China}\\
       \email{\{lyang, aiqy, croft\}@cs.umass.edu, guojiafeng@ict.ac.cn}
}

\maketitle

\begin{abstract}
As an alternative to question answering methods based on feature engineering, deep learning approaches such as convolutional neural networks (CNNs) and Long Short-Term Memory Models (LSTMs) have recently been proposed for semantic matching of questions and answers. To achieve good results, however, these models have been combined with additional features such as word overlap or BM25 scores. Without this combination, these models perform significantly worse than methods based on linguistic feature engineering. In this paper, we propose an attention based neural matching model for ranking short answer text.  We adopt value-shared weighting scheme instead of position-shared weighting scheme for combining different matching signals and incorporate question term importance learning using question attention network. Using the popular benchmark TREC QA data, we show that the relatively simple aNMM model can significantly outperform other neural network models that have been used for the question answering task, and is competitive with models that are combined with additional features. When aNMM is combined with additional features, it outperforms all baselines.
\end{abstract}

\ccsdesc[500]{Information systems~Retrieval models and ranking}
\ccsdesc[500]{Information systems~Question answering}


%
%
\printccsdesc


\keywords{Question Answering; Deep Learning; Value-shared Weights; Term Importance Learning }

\section{Introduction}
\label{sec:intro}

Question answering (QA), which returns exact answers as either short facts or long passages to natural language questions issued by users, is a challenging task and plays a central role in the next generation of advanced web search \cite{citeulike:9606736,Sun:2015:ODQ:2736277.2741651}.
Many of current QA systems use a learning to rank approach that encodes question/answer pairs with complex linguistic features including lexical, syntactic and semantic features \cite{Severyn:2015:LRS:2766462.2767738,Surdeanu08learningto}. For instance, Surdeanu et al. \cite{Surdeanu08learningto, Surdeanu:2011:LRA:2000517.2000520} investigated a wide range of feature types including similarity features, translation features, density/frequency features and web correlation features for learning to rank answers and show improvements in accuracy. However, such methods rely on manual feature engineering, which is often time-consuming and requires domain dependent expertise and experience. Moreover, they may need additional NLP parsers or external knowledge sources that may not be available for some languages.


Recently, researchers have been studying deep learning approaches to automatically learn semantic match between questions and answers. Such methods are built on the top of neural network models such as convolutional neural networks (CNNs) \cite{Yu:2014, Severyn:2015:LRS:2766462.2767738, QiuH15} and Long Short-Term Memory Models (LSTMs) \cite{DBLP:conf/acl/WangN15}. The proposed models have the benefit of not requiring hand-crafted linguistic features and external resources. Some of them \cite{Severyn:2015:LRS:2766462.2767738} achieve state-of-the art performance for the answer sentence selection task benchmarked by the TREC QA track. However, the weakness of the existing studies is that the proposed deep models, either based on CNNs or LSTMs,  need to be combined with additional features such as word overlap features and BM25 to perform well. Without combining these additional features, their performance is significantly worse than the results obtained by the state-of-the-art methods based on linguistic feature engineering \cite{yih-EtAl:2013:ACL2013}. This led us to propose the following research questions:

\textbf{RQ1} \textit{ Without combining additional features, could we build deep learning models that can achieve comparable or even better performance than methods using feature engineering ?}

\textbf{RQ2} \textit{ By combining additional features, could our model outperform state-of-the-art models for question answering ? }

To address these research questions, we analyze the existing current deep learning architectures for answer ranking and make the following two key observations:


\begin{enumerate}
	\item  \textbf{Architectures not specifically designed for question/answer matching:} Some methods employ CNNs for question/answer matching. However, CNNs are originally designed for computer vision (CV), which uses position-shared weights with local perceptive filters, to learn spatial regularities in many CV tasks. However, such spatial regularities may not exist in semantic matching between questions and answers, since important similarity signals between question and answer terms could appear in any position due to the complex linguistic property of natural languages. Meanwhile, models based on LSTMs view the question/answer matching problem in a sequential way. Without direct interactions between question and answer terms, the model may not be able to capture sufficiently detailed matching signals between them.
	
	\item  \textbf{Lack of modeling question focus:} Understanding the focus of questions, e.g., important terms in a question, is helpful for ranking the answers correctly . For example, given a question like ``\textit{Where was the first burger king restaurant opened ?}'', it is critical for the answer to talk about ``burger'', ``king'', ``open'', etc. Most existing text matching models do not explicitly model question focus. For example, models based on CNNs treat all the question terms as equally important when matching to answer terms. Models based on LSTMs usually model question terms closer to the end to be more important.
\end{enumerate}

To handle these issues in the existing deep learning architectures for ranking answers, we propose an attention based neural matching model (\textbf{aNMM}). The novel properties of the proposed model and our contributions can be summarized as follows:



\begin{enumerate}
	\item \textbf{Deep neural network with value-shared weights}:  We introduce a novel value-shared weighting scheme in deep neural networks as a counterpart of the position-shared weighting scheme in CNNs, based on the idea that semantic matching between a question and answer is mainly about the (semantic similarity) value regularities rather than spatial  
	regularities.
	\item \textbf{Incorporate attention scheme over question terms}: We incorporate the attention scheme over the question terms using a gating function, so that we can explicitly discriminate the question term importance.
	
	\item \textbf{Extensive experimental evaluation and promising results}. We perform a thorough experimental study based on the TREC QA dataset from TREC QA tracks 8-13, which appears to be one of the most widely used benchmarks for answer re-ranking. Unlike previous methods using CNNs \cite{Yu:2014, Severyn:2015:LRS:2766462.2767738} and LSTMs \cite{DBLP:conf/acl/WangN15}, which showed inferior results without combining additional features, our model can achieve better performance than a state-of-art method using linguistic feature engineering and comparable performance with previous deep learning models with combined additional features. If we combine our model with a simple additional feature like QL, our method can achieve the state-of-the-art performance among current existing methods for ranking answers under multiple metrics.
\end{enumerate}






\textbf{Roadmap}. The rest of our paper is organized as follows.  We will review related work in Section \ref{sec:rel}. In Section \ref{sec:aNMM_model_all}, we will present the proposed aNMM model with two components: value-shared weights and question attention network with gating functions. Two different architectures will be presented and analyzed. Section \ref{sec:exp} is a systematic experimental analysis using the TREC QA benchmark dataset. Finally, we conclude our paper and discuss future work in Section \ref{sec:conclu}. 

\section{Related Work}
\label{sec:rel}

Our work is related to several research areas, including deep learning models for text matching, factoid question answering, answer ranking in CQA and answer passage / sentence retrieval.

\textbf{Deep Learning Models for Text Matching.}
Recently there have been many deep learning models proposed for text matching and ranking. Such deep learning models include DSSM \cite{Huang:2013:LDS:2505515.2505665}, CDSSM \cite{DBLP:conf/emnlp/GaoPGHD14,DBLP:conf/cikm/ShenHGDM14}, ARC-I/ARC-II\cite{NIPS2014_5550} , DCNN \cite{KalchbrennerACL2014}, DeepMatch \cite{NIPS2013_5019}, MultiGranCNN \cite{yin-schutze:2015:ACL-IJCNLP} and MatchPyramid \cite{DBLP:conf/aaai/PangLGXWC16}. DSSM performs a non-linear projection to map the query and the documents to a common semantic space. The neural network models are trained using clickthrough data such that the conditional likelihood of the clicked document given the query is maximized. DeepMatch uses a topic model to construct the interactions between two texts and then makes different levels of abstractions with a deep architecture to model the relationships between topics.  ARC-I and ARC-II are two different architectures proposed by Hu et. al. \cite{NIPS2014_5550} for matching natural language sentences. ARC-I firstly finds the representation of each sentence and then compares the representations of the two sentences with a multi-layer perceptron (MLP). The drawback of ARC-I is that it defers the interaction between two sentences until their individual representation matures in the convolution model, and therefore has the risk of losing details, which could be important for the matching task. On the other hand, ARC-II is built directly on the interaction space between two sentences. Thus ARC-II makes two sentences meet before their own high-level representations mature, while still retaining the space for individual development of abstraction of each sentence. Our aNMM architecture adopts a similar design with ARC-II in the QA matching matrix where we build neural networks directly on the interaction of sentence term pairs. However, we adopt value-shared weights instead of position-shared weights as in the CNN used by ARC-II. We also add attention scheme to learn question term importance.

\textbf{Factoid Question Answering}. There have been many previous studies on factoid question answering, most of which use the benchmark data from TREC QA track \cite{yih-EtAl:2013:ACL2013,DBLP:conf/acl/WangN15,DBLP:conf/naacl/YaoDCC13,Yu:2014,Severyn:2015:LRS:2766462.2767738}. Yih et. al. \cite{yih-EtAl:2013:ACL2013} formulated answer sentence selection as a semantic matching problem with a latent word-alignment structure and conducted a series of experimental studies on leveraging proposed lexical semantic models. Iyyer et. al. \cite{Iyyer:Boyd-Graber:Claudino:Socher:Daume-2014} introduced a recursive neural network (RNN) model that can reason over text that contains very few individual words by modeling textual compositionality. Yu~et~al.~\cite{Yu:2014} proposed an approach for answer sentence selection via distributed representations, and learned to match questions with answers by considering their semantic encoding. They combined the learning results of their model with word overlap features by training a logistic regression classifier.  Wang and Nyberg \cite{DBLP:conf/acl/WangN15} proposed a method which uses a stacked bidirectional Long-Short Term Memory (BLSTM) network to sequentially read words from question and answer sentences, and then output their relevance scores. Their system needs to combine the stacked BLSTM relevance model with a BM25 score to achieve good performance. Severyn and Moschitti \cite{Severyn:2015:LRS:2766462.2767738} presented a convolutional neural network architecture for re-ranking pairs of short texts, where they learned the optimal representation of text pairs and a similarity function to relate them in a supervised way from the available training data. They also need to combine additional features into their model to outperform previous methods. Unlike the previous research, our method can outperform previous methods using feature engineering without combining any additional features. With an additional simple feature like QL, our model is significantly better than the previous state-of-the-art methods including deep learning methods.

\textbf{Answer Ranking in CQA}. There is also previous research on ranking answers from community question answering (CQA) sites. Surdeanu~et~al.~\cite{Surdeanu08learningto, Surdeanu:2011:LRA:2000517.2000520} investigated a wide range of feature types such as similarity features, translation features, density / frequency features for ranking answers to non-factoid questions in Yahoo! Answers. Jansen et al.~\cite{jansen_discourse_2014} presented an answer re-ranking model for non-factoid questions that integrate lexical semantics with discourse information driven by two representations of discourse. Xue et al.~\cite{xue_retrieval_2008} proposed a retrieval model that combines a translation-based language model for the question part with a query likelihood approach for the answer part. Questions from CQA sites are mostly non-factoid questions. Our research is closer to  factoid questions such as questions in TREC QA data.

\begin{figure*}[th]
	\center
	\includegraphics*[viewport=0mm 0mm 190mm 105mm, scale=0.80]{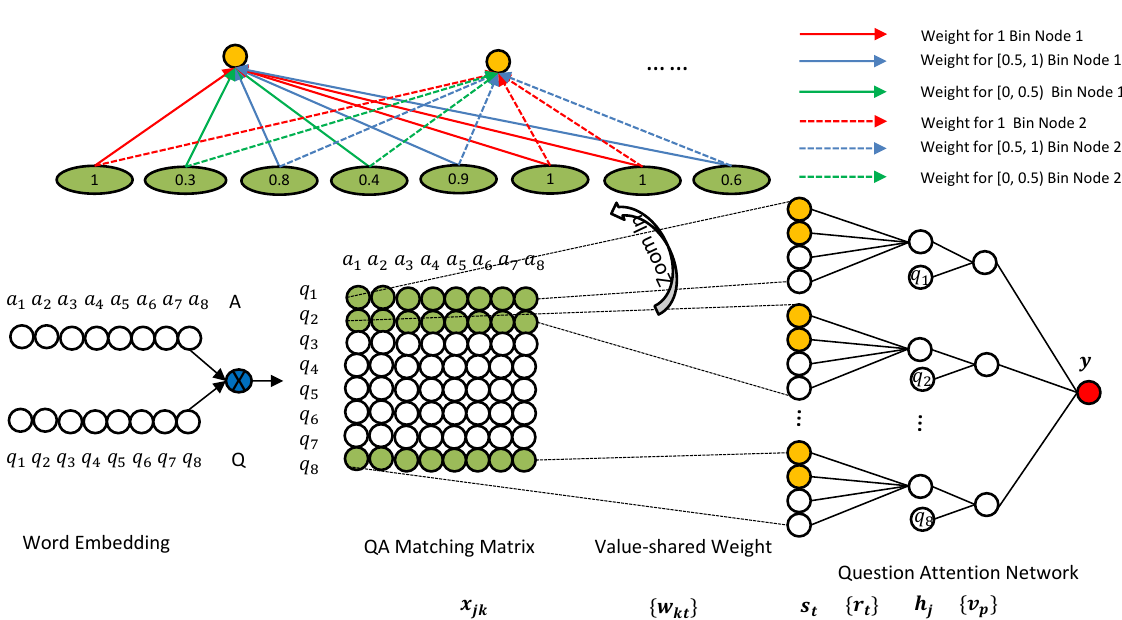}
	\vspace{-0.3cm}
	\caption{The proposed architecture of attention-based neural matching model (aNMM-2) for ranking answers.}\label{fig:gnnm2V2}
\end{figure*}

\textbf{Answer Passage / Sentence Retrieval}. Our work is also related to previous research on answer passage / sentence retrieval. Tymoshenko and Moschitti \cite{Tymoshenko:2015:AIS:2806416.2806490} studied the use of syntactic and semantic structures obtained with shallow and deeper syntactic parsers in the answer passage re-ranking task. Keikha et al.~\cite{keikha_retrieving_2014,Keikha:2014:EAP:2600428.2609485} developed an annotated data set for non-factoid answer finding using TREC GOV2 collections and topics. They annotated passage-level answers, revisited several passage retrieval models with this data, and came to the conclusion that the current methods are not effective for this task. Yang et al. \cite{DBLP:conf/ecir/YangASCPCGS16} developed effective methods for answer sentence retrieval using this annotated data by combining semantic features, context features and basic text matching features with a learning to rank approach. Our model is built on attention-based neural matching model with value-shared weighting schema. Unlike learning to rank approaches with feature engineering, our model can achieve good performance for ranking answers without any additional manual feature engineering, preprocessing of NLP parsers and external resources like knowledge bases.



\section{Attention-based Neural Matching Model}
\label{sec:aNMM_model_all}

In this section we present the proposed model referred as \textbf{aNMM} (attention-based Neural Matching Model), which is shown in Figure \ref{fig:gnnm2V2}. Before we introduce our model, we firstly define some terminologies.

\subsection{Terminology}
\label{sec:terminology}

\noindent \textbf{Short Answer Text:} We use \textit{Short Answer Text} to refer to a short fact, answer sentences or answer passages that can address users' information needs in the issued questions. This is the ranking object in this paper and includes answers in various lengths. In the experiments of this paper, we mainly focus on ranking answer sentences that contain correct answer facts as in TREC QA data.

\noindent \textbf{QA Matching Matrix:} We use \textit{QA Matching Matrix} to refer to a matrix which represents the semantic matching information of term pairs from a question and answer pair. Given a question $\mathbf{q}$ with length $M$ and an answer $\mathbf{a}$ with length $N$, a QA matching matrix is an $M$ by $N$ matrix $\mathbf{P}$, where $\mathbf{P}_{j,i}$ denote the semantic similarity between term $\mathbf{q}_j$ and term $\mathbf{a}_i$ measured by the cosine similarity of the corresponding word embeddings of terms. If  $\mathbf{q}_j$ and $\mathbf{a}_i$ are the same term, we assign $\mathbf{P}_{j,i}$ as $1$.

\noindent \textbf{QA Matching Vector:} We use \textit{QA Matching Vector} to refer to a row in the QA matching matrix. As presented before, the $j$-th row of the QA matching matrix $\mathbf{P}$ contains the semantic similarity between $\mathbf{q}_j$ and all terms in answer $\mathbf{a}$ . 

 \begin{figure*}[th]
 	\center
 	\includegraphics[width=6.9in, height=1.9in]{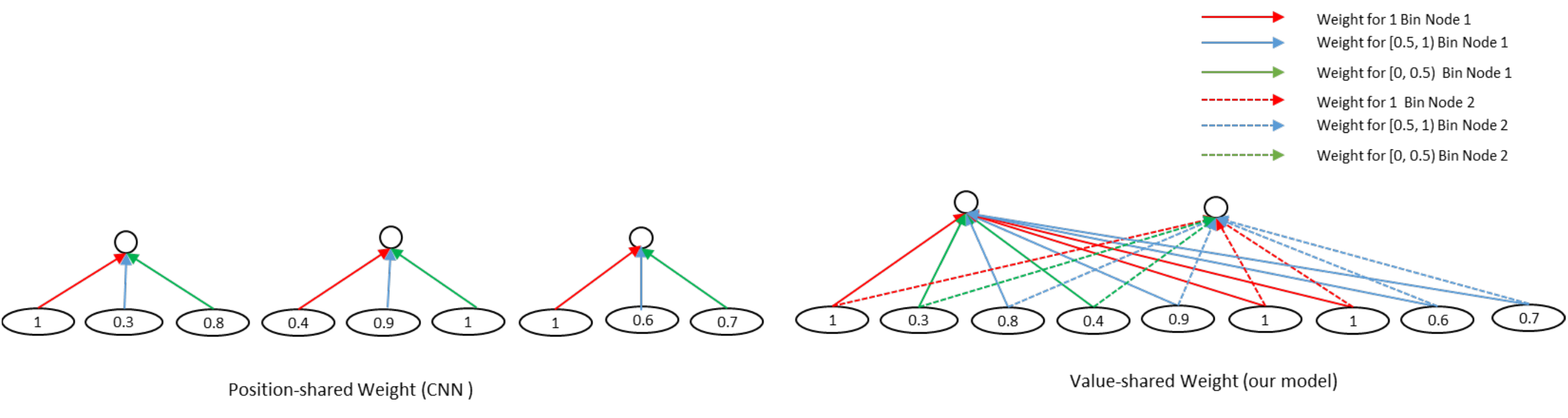}\\
 	\vspace{-0.4cm}
 	\caption{The comparison of position-shared weight in CNN and value-shared weight in aNMM. In CNN, the weight associated with a node only depends on its position or relative location as specified by the filters. In aNMM, the weight associated with a node depends on its value.}\label{fig:PTWVTW}
 	\vspace{-0.4cm}
 \end{figure*}

\subsection{Model Overview}
\label{sec:method_overview}

Our method contains three steps as follows:

\begin{enumerate}
	\item We construct QA matching matrix for each question and answer pair with pre-trained word embeddings.
	\item We then employ a deep neural network with value-shared weighting scheme in the first layer, and fully connected layers in the rest to learn hierarchical abstraction of the semantic matching between question and answer terms.
	\item  Finally, we employ a question attention network to learn question term importance and produce the final ranking score.
\end{enumerate}

 We propose two neural matching model architectures and compare the effectivenesses of them. We firstly describe a basic version of the architecture, which is referred to as \textbf{aNMM-1}. 
 
 In the following sections, we will explain in detail the two major designs of aNMM-1, i.e., value-shared weights and question attention network.

\subsection{Value-shared Weighting}
\label{sec:binweight}

 We first train word embeddings with the Word2Vec tool by Mikolov et al.\cite{NIPS2013_5021} with the English Wikipedia dump to construct QA matching matrices. Given a question sentence and an answer sentence, we compute the dot product of the normalized word embeddings of all term pairs to construct the QA matching matrix $\mathbf{P}$ as defined in Section \ref{sec:terminology}. A major problem with the QA matching matrix is the variable size due to the different lengths of answers for a given question. To solve this problem, one can use CNN with pooling strategy to handle the variable size. However, as we have mentioned before, CNNs basically use position-shared weighting scheme which may not fit semantic matching between questions and answers. Important question terms and semantically similar answer words could appear anywhere in questions/answers due to the complex linguistic property of natural languages. Thus we adopt the following method to handle the various length problem:
 
 \textit{Value-shared Weights}: For this method, the assumption is that matching signals in different ranges play different roles in deciding the final ranking score. Thus we introduce the value-shared weighting scheme to learn the importance of different levels of matching signals. The comparison between the position-shared weight and value-shared weight is shown in Figure \ref{fig:PTWVTW}. We can see that for position-shared weights, the  weight associated with a node only depends on its position or relative location as specified by the filters in CNN. However in our model, the weight associated with a node depends on its value. The value of a node denotes the strength of the matching signal between term pairs of questions and answers from the QA matching matrix, as explained in Section \ref{sec:terminology}. Such a setting enables us to use the learned weights to encode how to combine different levels of matching signals. After this step, the size of the hidden representation becomes fixed and we can use normal fully connected layers to learn higher level representations. We use the term \textit{bin} to denote a specific range of matching signals. since $\mathbf{P}_{j,i} \in [-1,1]$, if we set the size of bins as $0.1$, then we have $21$ bins where there is a separate bin for $\mathbf{P}_{j,i} = 1$ to denote exact match of terms.
 
 Specifically, value-shared weights are adopted in the forward propagation prediction process from the input layer to the hidden layer over each question term in aNMM-1 as follows:
 
 \textbf{Input Layer to Hidden Layer}.
 Let $\mathbf{w}$ denote a $K+1$ dimensional model parameter from input layer to hidden layer. $x_{jk}$ denotes the sum of all matching signals within the $k$-th value range or bin. For each QA matching vector of a given query $\mathbf{q}$, the combined score after the activation function of the $j$-th node in hidden layer is defined as 
 
 
 \begin{eqnarray}
 h_j = \delta (\sum_{k=0}^{K}w_k \cdot x_{jk})
 \end{eqnarray}
 
 where $j$ is the index of question words in $\mathbf{q}$. We use the sigmoid function as the activation function, which is commonly adopted in many neural network architectures.

 
 
 

\subsection{Question Attention Network}

In addition to value-shared weighting, another model component of aNMM-1 is the question attention network. In a committee of neural networks which  consists of multiple networks, we need to combine the output of these networks to output a final decision vector. The question attention network uses the gating function \cite{suGatingNN} to control the output of each network in this process. Specifically, in aNMM-1 we use the softmax gate function to combine the output of multiple networks where each network corresponds to a question term as shown in Figure \ref{fig:gnnm2V2}. We feed the dot product of query word embedding and model parameter to the softmax function to represent the query term importance. In this setting, we can directly compare the relative term importance of query words within the same query with softmax function. We also tried sigmoid gate function, but this did not perform as well as softmax gate function.

Softmax gate function is used in the forward propagation process from the hidden layer to the output layer as follows:

\textbf{Hidden Layer to Output Layer}.
From the hidden layer to the output layer, we add a softmax gate function to learn question attention. Let $\mathbf{v}$ denote a $P$ dimensional vector which is a model parameter. We feed the dot product of query word embedding $\mathbf{q}_j$ and $\mathbf{v}$ to the softmax function to represent the query term importance as shown in Equation \ref{Eqn:forward_softmax_gate}. Note that we normalize the query word embedding before computing the dot product.

\begin{eqnarray} \label{Eqn:forward_softmax_gate}
y = \sum_{j=1}^{M} g_j \cdot h_j = \sum_{j=1}^{M} \frac{\exp({\mathbf{v}\cdot \mathbf{q}_j)} }{\sum_{l=1}^{L}\exp({\mathbf{v}\cdot \mathbf{q}_l})}  \cdot \delta (\sum_{k=0}^{K}w_k \cdot x_{jk})
\end{eqnarray}


Unlike previous models like CNNs \cite{Severyn:2015:LRS:2766462.2767738}  and BLSTM \cite{DBLP:conf/acl/WangN15}, which learn the semantic match score between questions and answers through representation learning from matching matrix or question / answer pair  sequences, aNMM achieves this by combining semantic matching signals of term pairs in questions and answers weighted by the output of question attention network, where softmax gate functions help discriminate the term importance or attention on different question terms.

\subsection{Model Training}


%

%

For aNMM-1, the model parameters contain two sets: 1) The value-shared weights $\mathbf{w_k}$ for combining matching signals from the input layer to the hidden layer. 2) The parameters $\mathbf{v_p}$ in the gating function from the hidden layer to the output layer. 

To learn the model parameters from the training data, we adopt a pair-wise learning strategy with a large margin objective. Firstly we construct triples $(\mathbf{q}, \mathbf{a}^+, \mathbf{a}^-)$ from the training data, with $\mathbf{q}$ matched with $\mathbf{a}^+$ better than with $\mathbf{a}^-$. We have the ranking-based loss as the objective function as following:


\begin{eqnarray} \label{eqn:pair_wise_loss_function}
e(\mathbf{q}, \mathbf{a}^+, \mathbf{a}^- ; \mathbf{w}, \mathbf{v}) = \max (0, 1 - S(\mathbf{q}, \mathbf{a}^+) + S(\mathbf{q},\mathbf{a}^-))
\end{eqnarray}

where $S(\mathbf{q}, \mathbf{a})$ denote the predicted matching score for QA pair $(\mathbf{q}, \mathbf{a})$. During training stage, we will scan all the triples in training data. Given a triple $(\mathbf{q}, \mathbf{a}^+, \mathbf{a}^-)$, we will compute $\Delta S = 1 - S(\mathbf{q}, \mathbf{a}^+) + S(\mathbf{q},\mathbf{a}^-) $. If $\Delta S \leq 0 $, we will skip this triple. Otherwise, we need to update model parameters with back propagation algorithm to minimize the objective function. 

Under softmax gate function setting, the gradients of $e$ w.r.t. $\mathbf{v}$ from hidden layer to the output layer is shown in Equation \ref{Eqn:grad_vp_softmax}

\begin{eqnarray} \label{Eqn:grad_vp_softmax}
\frac{\partial e}{\partial v_p} = \sum_{j=1}^{M} \frac{\partial g_j}{\partial v_p} \cdot (- \delta (u^+) + \delta (u^-))
\end{eqnarray}

where 

\begin{eqnarray}
u^+ = \sum_{k=0}^{K}w_k \cdot x^+_{jk}, u^- = \sum_{k=0}^{K}w_k \cdot x^-_{jk} \nonumber
\end{eqnarray}
	
$\frac{\partial g_j}{\partial v_p}$ can be derived as 


\begin{scriptsize}
	\begin{eqnarray} \label{Eqn:parital_gj_vp}
	\frac{\exp (\mathbf{v}\cdot \mathbf{q}_j)\cdot q_{jp} \sum_{l=1}^{M}\exp(\mathbf{v}\cdot \mathbf{q}_l) - \exp (\mathbf{v}\cdot \mathbf{q}_j)\sum_{l=1}^{M}\exp(\mathbf{v}\cdot \mathbf{q}_l) \cdot q_{lp}}{ ( \sum_{l=1}^{M} \exp(\mathbf{v} \cdot \mathbf{q}_l))^2}  \nonumber
	\end{eqnarray}
\end{scriptsize}

The gradient of $e$ w.r.t. $\mathbf{w}$ from input layer to hidden layer is shown in Equation \ref{Eqn:grad_wk_softmax}.

\begin{eqnarray} \label{Eqn:grad_wk_softmax}
\frac{\partial e}{w_k} &=& \sum_{j=1}^{M}\frac{\exp({\mathbf{v}\cdot \mathbf{q}_j)} }{\sum_{l=1}^{L}\exp({\mathbf{v}\cdot\mathbf{q}_l})} \cdot (-\delta(u^+)(1-\delta(u^+)) x^+_{jk} \nonumber \\
&& + \delta(u^-)(1-\delta(u^-)) x^-_{jk})
\end{eqnarray}

With the formulas of gradients, we can perform stochastic gradient descent to learn model parameters. We use mini-batch gradient descent to achieve more robust performance on the ranking task. For the learning rate, we adopt adaptive learning rate: $\eta = \eta_0 (1-\epsilon)$, where $\epsilon$ will approach $1$ with more iterations. Such a setting has better guarantee for convergence.

\begin{figure*}[th]
	\centering
	\includegraphics[width=\textwidth]{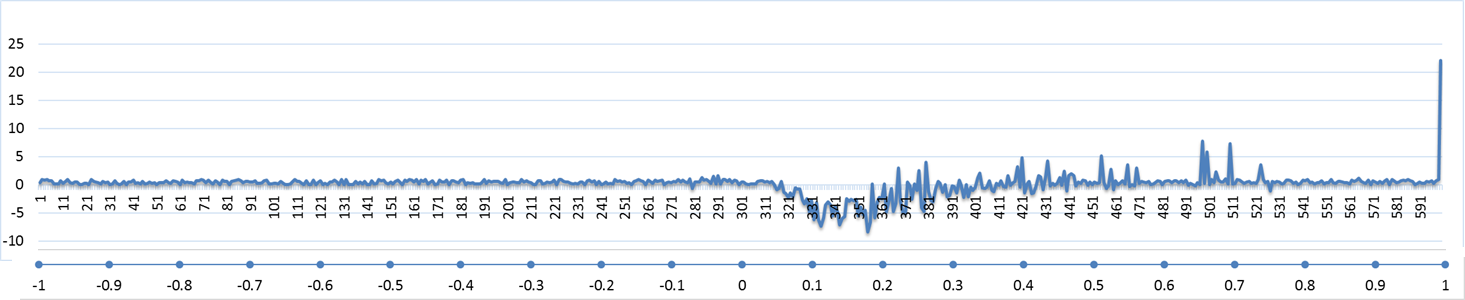}
	\caption{Visualization of learned value-shared weigths of aNMM-1. The x-axis is index of bin ranges and the y-axis is the value-shared weights corresponding to each bin range. The range of match signals is [-1,1] from the left to the right.}
	\label{fig:vtw_anmm}
\end{figure*}


\subsection{Extension to Deep Neural Networks with Multiple Sets of Value-shared Weights}
\label{ext_multi_layer_aNMM-2}

In aNMM-1, we can only use one set of value-shared weights for each QA matching vector. We further propose a more flexible neural network architecture which could enable us to use multiple sets of value-shared weights for each QA matching vector, leading to multiple intermediate nodes in the first hidden layer, as shown in Figure \ref{fig:gnnm2V2} by the yellow color. We refer to this extended model as \textbf{aNMM-2}.  The model architecture shown in Figure \ref{fig:gnnm2V2} is corresponding to aNMM-2.

\subsubsection{Forward Propagation Prediction}
For aNMM-2, we add a hidden layer in the neural network where we learn multiple combined scores from the input layer. With this hidden layer, we define multiple weight vectors as $\mathbf{w}$. Thus $\mathbf{w}$ becomes a two dimensional matrix. The formula for the forward propagation prediction is as follows:

\begin{equation} \label{eqn:gnn2_forward_prediction}
y = \sum_{j=1}^{M} \tau (\mathbf{v} \cdot \mathbf{q}_j)  \cdot \delta(\sum_{t=0}^{T}r_t \cdot \delta (\sum_{k=0}^{K} w_{kt} x_{jk} ))
\end{equation}

where  $ \tau (\mathbf{v} \cdot \mathbf{q}_j) = \frac{\exp({\mathbf{v}\cdot \mathbf{q}_j)} }{\sum_{l=1}^{L}\exp({\mathbf{v}\cdot\mathbf{q}_l})}$ and $\tau$ denote the softmax gate function. 
$T$ is the number of nodes in hidden layer 1. $r_t$ is the model parameter from hidden layer 1 to hidden layer 2, where we feed the linear combination of outputs of nodes in hidden layer 1 to an extra activation function comparing with Equation \ref{Eqn:forward_softmax_gate}. Then from hidden layer 2 to output layer, we sum over all outputs of nodes in hidden layer 2 weighted by the outputs of softmax gate functions, which also form the question attention network.

\subsubsection{Back Propagation for Model Training}
For aNMM-2, we have three sets of model parameters: 1) $w_{kt}$ from input layer to hidden layer 1; 2) $r_t$ from hidden layer 1 to hidden layer 2; 3) $v_p$ from hidden layer 2 to output layer. All three sets of parameters are updated through back propagation. The definition of the objective function is the same as Equation \ref{eqn:pair_wise_loss_function}. The back propagation process for model parameter learning is described as follows:

\textbf{From hidden layer 2 to output layer.}
The gradients of the objective function w.r.t. $\mathbf{v}$ is as following:

	\begin{eqnarray} \label{Eqn:grad_vp_softmax_gnnm2}
	\frac{\partial e}{\partial v_p} = \sum_{j=1}^{M} \frac{\partial g_j}{\partial v_p} \cdot (- h_j^+ + h_j^-)
	\end{eqnarray}

Where 

$h_j^+ = \delta(\sum_{t=0}^{T} r_t \cdot \delta (\sum_{k=0}^{K} w_{kt}x_{jk}^+))$

$h_j^- = \delta(\sum_{t=0}^{T} r_t \cdot \delta (\sum_{k=0}^{K} w_{kt}x_{jk}^-))$


\textbf{From hidden layer 1 to hidden layer 2.}
The gradients of the objective function w.r.t. $\mathbf{r}$ is as following:

\begin{eqnarray}
\frac{\partial e}{\partial r_t} &=& \sum_{j=1}^{M} \tau (\mathbf{v} \cdot \mathbf{q}_j) ( - h_j^+ )(1- h_j^+   )s_t^+  + h_j^-   (1- h_j^-  )s_t^-) \nonumber
\end{eqnarray}

Where 



$s_t^+ = \delta(\sum_{k=0}^{K} w_{kt}x_{jk}^+ )$

$s_t^- = \delta(\sum_{k=0}^{K} w_{kt}x_{jk}^- )$.

\textbf{From input layer to hidden layer 1.}

The gradients of the objective function w.r.t. $\mathbf{w}$ is as following:

\begin{eqnarray}
	\frac{\partial e}{\partial w_{kt}} &=& \sum_{j=1}^{M} \tau (\mathbf{v} \cdot \mathbf{q}_j) (- \frac{\partial y^+}{u_j^{'+}} \cdot r_t \cdot \delta(u_t^+)(1- \delta (u_t^+)) \cdot x^+_{jk} \nonumber \\ 
	&& +   \frac{\partial y^-}{u_j^{'-}} \cdot r_t \cdot \delta(u_t^-)(1- \delta (u_t^-)) \cdot x^-_{jk}   )
\end{eqnarray}

Where

$u_j^{'+} = \sum_{t=0}^{T} r_t \cdot \delta (\sum_{k=0}^{K} w_{kt}x_{jk}^+)$

$u_j^{'-} = \sum_{t=0}^{T} r_t \cdot \delta (\sum_{k=0}^{K} w_{kt}x_{jk}^-)$


Initially we will randomly give the values of model parameters. Then we will use back propagation to update the model parameters. When the learning process converge, we use the learned model parameters for prediction to rank short answer texts. 

%
%
%
\section{Experiments}
\label{sec:exp}

 \begin{table*}[t]
 	\centering
 	\caption{The statistics of the TREC QA data set.}
 	\begin{tabular}{ l || l | l  | l | l | l    } 
 		\hline
 		Data & \#Questions & \#QA pairs & \%Correct & \#Answers/Q & Judgement\\\hline
 		TRAIN-ALL & 1,229 & 53,417 & 12.00\% & 43.46 & automatic \\
 		TRAIN & 94 & 4,718 & 7.40\% & 50.19 & manual \\
 		DEV & 82 & 1,148 & 19.30\% & 14.00 & manual \\
 		TEST & 100 & 1,517 & 18.70\% & 15.17 & manual \\\hline
 	\end{tabular}
 	\label{tab:trecqa_data}
 \end{table*}

 \begin{table*}[t]
 	\centering
 	\caption{Examples of learned question term importance by aNMM-1.}
 	\begin{tabular}{ l || l | l  | l | l | l | l | l | l  } 
 		\hline
 		test\_14 & when & did & the & khmer & rouge & come & into & power \\
 		Term Importance & 4.91E-03 & 7.18E-04 & 8.97E-04 & 5.67E-01 & 2.13E-01 & 1.81E-02 & 6.59E-03 & 1.89E-01 \\ \hline
 		test\_66 & where & was & the & first & burger & king & restaurant & opened \\
 		Term Importance & 2.16E-04 & 5.67E-04 & 1.96E-04 & 2.57E-03 & 3.43E-01 & 4.39E-01 & 5.35E-03 & 2.08E-01 \\ \hline
 		train\_84 & at & what & age & did & rossini & stop & writing & opera \\
 		Term Importance & 5.06E-02 & 2.54E-03 & 6.17E-02 & 2.68E-03 & 3.89E-01 & 4.28E-01 & 9.29E-03 & 5.64E-02\\ \hline
 		
 	\end{tabular}
 	\label{tab:query_import}
 \end{table*}

 \begin{figure*}[th]
 	\center
 	\includegraphics[width=6.2in, height=1.75in]{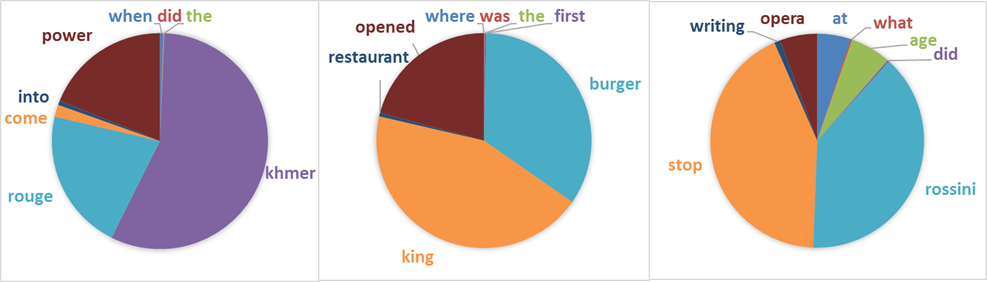}\\
 	\vspace{-0.1cm}
 	\caption{Visualization of learned question term importance by aNMM-1.}\label{fig:queryTermImport}
 	\vspace{-0.4cm}
 \end{figure*}
 
\subsection{Data Set and Experiment Settings}
\label{sec:data_and_settings}
We use  the TREC QA data set \footnote{\url{https://github.com/aseveryn/deep-qa}} created by Wang et. al. \cite{wang-smith-mitamura:2007:EMNLP-CoNLL2007} from TREC QA track 8-13 data, with candidate answers automatically selected from each question's document pool using a combination of overlapping non-stop word counts and pattern matching. This data set is one of the most widely used benchmarks for answer re-ranking. Table \ref{tab:trecqa_data} shows the statistics of this data set. The dataset contains a set of factoid questions with candidate answers which are limited to a single sentence.  There are two training data sets: TRAIN and TRAIN-ALL. Answers in TRAIN have manual judgments for the answer correctness. The manual judgment of candidate answer sentences is provided for the entire TREC 13 set and for a part of questions from TREC 8-12. TRAIN-ALL is another training set with much larger number of questions. The correctness of candidate answer sentences in TRAIN-ALL is identified by matching answer sentences with answer pattern regular expressions provided by TREC. This data set is more noisy, however it provides many more QA pairs for model training. There is a DEV set for hyper-parameter optimization and TEST set for model testing. We use the same train/dev/test partition in our experiments to directly compare our results with previous works. For data preprocess, we perform tokenization without stemming. We maintain stop words during the model training stage.

\textbf{Word Embeddings}. We obtain pre-trained word embeddings with the Word2Vec tool by  Mikolov et al.\cite{NIPS2013_5021} with the English Wikipedia dump.  We use the skip-gram model with window size $5$ and filter words with frequency less than $5$ following the common practice in many neural embedding models. For the word vector dimension, we tune it as a hyper-parameter on the validation data starting from $200$ to $1000$. 
Embeddings for words not present are randomly initialized with sampled numbers from uniform distribution U[-0.25,0.25], which follows the same setting as \cite{Severyn:2015:LRS:2766462.2767738}. 

\textbf{Model Hyper-parameters}. For the setting of hyper-parameters, we set the number of bins as $600$, word embedding dimension as $700$ for aNNM-1, the number of bins as $200$, word embedding dimension as $700$ for aNNM-2 after we tune hyper-parameters on the provided DEV set of TREC QA data.

\subsection{Evaluation and Metrics}
For evaluation, we rank answer sentences with the predicted score of each method and compare the rank list with the ground truth to compute metrics. We choose Mean Average Precision (MAP) and Mean Reciprocal Rank (MRR), which are commonly used in information retrieval and question answering, as the metric to evaluate our model.

The definition of MRR is as follows:

$MRR = \frac{1}{|Q|} \sum_{q \in Q} \frac{1}{rank(fa)}$

where $rank(fa)$ is the position of the first correct answer in the rank list for the question $q$. Thus MRR is only based on the rank of the first correct answer. It is more suitable for the cases where the rank of the first correct answer is emphasized or each question only have one correct answer. On the other hand, MAP looks at the ranks of all correct answers. It is computed as following:

$MAP = \frac{1}{|Q|} \sum_{q \in Q} AP(q)$

where $AP(q)$ is the average precision for each query $q \in \mathbf{Q}$. Thus MAP is the average performance on all correct answers. We use the official $trec\_eval$\footnote{\url{http://trec.nist.gov/trec_eval/}} scripts for computing these metrics.

\subsection{Model Learning Results}
In this section, we give some qualitative analysis and visualization of our model learning results. Specifically, we analyze the learned value-shard weights and question term importance by aNMM.

\subsubsection{Value-shared Weight}
We take the learned value-shared weights of aNMM-1 as the example. Figure \ref{fig:vtw_anmm} shows the learned value-shared weights by aNMM-1. In aNMM-1, for each QA matching vector, there is only one bin node. Thus the learned value-shared weights for aNMM-1 is a one dimension vector. For aNMM-1, we set the number of bins as $600$ as presented in Section \ref{sec:data_and_settings}. Note that the x-axis is the index of bin range and the y-axis is the value-shared weights corresponding to each bin range. The range of match signals is [-1,1] from the left to the right. We make the following observations: (1) The exact match signal which is corresponding to $1$ in the last bin is tied with a very large weight, which shows that exact match information is very important. (2) For positive matching score from $(0,1)$, which is corresponding to bin index $(300,600)$, the learned value-shared weights are different for matching score range $(0.5,1)$ (bin index $(450,600)$) and matching score range $(0,0.5)$ (bin index $(300,450)$) . We can observe many positive value-shared weights for matching score range$(0.5,1)$  and negative value-shared weights for matching score range$(0,0.5)$. This makes sense since high semantic matching scores are positive indicators on answer correctness, whereas low semantic matching scores indicate that the candidate answer sentences contain irrelevant terms. (3) For negative matching scores from $(-1,0)$, we can see there is not a lot of differences between value-shared weights for different ranges. A major reason is that most similarity scores based on word embeddings are positive. Therefore, we can remove bins corresponding to negative matching scores to reduce the dimension of value-shared weight vectors, which can help improve the efficiency of the model training process. We will show more quantitative results on the comparison between value-shared weights and position-shared weights in CNN in Section \ref{sec:exp_res_rank_answers}.

\subsubsection{Question Term Importance}
\label{sec:QueryImport}
Next we analyze the learned question term importance of our model. Due to the space limit, we also use the learned question term importance of aNMM-1 as an example. Table \ref{tab:query_import} shows the examples of learned question term importance by aNMM-1. We also visualize the question term importance in Figure \ref{fig:queryTermImport}. Based on the results in the table and the figure, we can clearly see that aNMM-1 learns reasonable term importance. For instance, with the question attention network, aNMM-1 discovers important terms like ``khmer'', ``rouge'', ``power'' as for the question ``\textit{When did the khmer rouge come into power ?}''. Terms like ``age'', ``rossinin'', ``stop'', ``writing'',``opera'' are highlighted for the question ``\textit{At what age did rossini stop writing opera ?} ''. For the question ``\textit{Where was the first burger king restaurant opened ?}'' mentioned in Section \ref{sec:intro}, ``burger'', ``king'', ``opened'' are treated as important question terms.


 \begin{table}[t]
 	\centering
 	\caption{The comparision of aNMM-1/aNMM-2 with aNMM-IDF which is a degenerate version of our model where we use IDF to directly replace the output of question attention network.}
 	\begin{tabular}{ l || l | l  || l | l   } 
 		\hline
 		\multicolumn{1}{l||}{Training Data } & \multicolumn{2}{c||}{TRAIN} & \multicolumn{2}{c}{TRAIN-ALL}\\ \hline 
 		Method & MAP & MRR  & MAP & MRR  \\\hline
 		aNMM-IDF & 0.6624 & 0.7376 & 0.7225 & 0.7873 \\\hline
 		aNMM-2 & 0.7191 & 0.7974 &\textbf{ 0.7407} & 0.7969 \\
 		aNMM-1 & \textbf{0.7334} & \textbf{0.8020} & 0.7385 &\textbf{0.7995} \\\hline
 	\end{tabular}
 	\label{tab:trecqa_res_comp_idfweight}
 \end{table}
 
An interesting question is how the learned term importance compare with traditional IR term weighting methods such as IDF. We design an experiment to compare aNMM-1/aNMM-2 with aNMM-IDF, which is a degenerate version of our model where we use IDF to directly replace the output of question attention network. In this case, $\tau (\mathbf{v} \cdot \mathbf{q}_j)$ in Equation \ref{eqn:gnn2_forward_prediction} is replaced by the IDF of the j-th question term. Table \ref{tab:trecqa_res_comp_idfweight} shows the results. We find that if we replace the output of question attention network of aNMM with IDF, it will decrease the answer ranking performance, especially on TRAIN data. Thus, we can see that with the optimization process in the back propagation training process, aNMM can learn better question term weighting score than heuristic term weighting functions like IDF.




\subsection{Experimental Results for Ranking Answers}
\label{sec:exp_res_rank_answers}

\subsubsection{Learning without Combining Additional Features}
Our first experimental setting is ranking answer sentences directly by the predicted score from aNMM without combining any additional features. This will enable us to answer RQ1 proposed in Section \ref{sec:intro}. Table \ref{tab:trecqares_withno_af_dl} shows the results of TREC QA on TRAIN and TRAIN-ALL without combining additional features. In this table, we compare the results of aNMM with other previous deep learning methods including  CNN \cite{Yu:2014, Severyn:2015:LRS:2766462.2767738} and LSTM \cite{DBLP:conf/acl/WangN15}. We summarize our observations as follows: (1) Both aNMM-1 and aNMM-2 show significant improvements for MAP and MRR on TRAIN and TRAIN-ALL data sets comparing with previous deep learning methods. Specifically, if we compare the results of aNMM-1 with the strongest deep learning baseline method by Severyn et al. \cite{Severyn:2015:LRS:2766462.2767738} based on CNN, we can see aNMM-1 outperform CNN for $14.67$\% in MAP on TRAIN, $9.15$\% in MAP on TRAIN-ALL. For MRR, we can also observe similar significant improvements of aNMM-1. These results show that with the value-shared weight scheme instead of the position-shared weight scheme in CNN and term importance learning with question attention network, aNMM can predict ranking scores with much higher accuracy comparing with previous deep learning models for ranking answers. (2) If we compare the results of aNMM-1 and aNMM-2, we can see their results are very close. aNMM-1 has slightly better performance than aNMM-2. This result indicates that adding one more hidden layer to incorporate multiple bin nodes does not necessarily increase the performance for answer ranking in TREC QA data. From the perspective of model efficiency, aNMM-1 could be a better choice since it can be trained much faster with good prediction accuracy. However, for larger training data sets than TREC QA data, aNMM-2 could have better performance since it has more model parameters and is suitable for fitting larger training data set. We leave the study of impact of the number of hidden layers in aNMM to future work.

 \begin{table}[t]
 	\centering
 	\caption{Results of TREC QA on TRAIN and TRAIN-ALL without combining additional features (Compare with deep learning methods).}
 	\begin{tabular}{ l || l | l  || l | l   } 
 		\hline
 		\multicolumn{1}{l||}{Training Data } & \multicolumn{2}{c||}{TRAIN} & \multicolumn{2}{c}{TRAIN-ALL}\\ \hline 
 		Method & MAP & MRR  & MAP & MRR  \\\hline
 		Yu et al. (2014) \cite{Yu:2014} & 0.5476 & 0.6437 & 0.5693 & 0.6613 \\
 		Wang et al.(2015) \cite{DBLP:conf/acl/WangN15}  & / & / & 0.5928 & 0.6721 \\
 		Severyn et al. (2015) \cite{Severyn:2015:LRS:2766462.2767738} & 0.6258 & 0.6591 & 0.6709 & 0.7280 \\\hline
 		aNMM-2 & 0.7191 & 0.7974 & \textbf{0.7407} & 0.7969 \\
 		aNMM-1 & \textbf{0.7334} & \textbf{0.8020} & 0.7385 & \textbf{0.7995} \\\hline
 	\end{tabular}
 	\label{tab:trecqares_withno_af_dl}
 \end{table}
 
 \begin{table}[t]
 	\centering
 	\caption{Results of TREC QA on TRAIN-ALL without combining additional features (Compare with methods using feature engineering).}
 	\begin{tabular}{ l || l | l     } 
 		\hline
 		Method & MAP & MRR  \\\hline
 		Wang et al. (2007) \cite{wang-smith-mitamura:2007:EMNLP-CoNLL2007} & 0.6029 & 0.6852  \\
 		Heilman and Smith (2010) \cite{Heilman:2010:TEM:1857999.1858143} & 0.6091 & 0.6917  \\
 		Wang and Manning (2010) \cite{Wang:2010:PTM:1873781.1873912} & 0.5951 & 0.6951  \\
 		Yao et al. (2013) \cite{DBLP:conf/naacl/YaoDCC13} & 0.6307 & 0.7477  \\
 		Severyn et al. (2013) \cite{DBLP:conf/emnlp/SeverynM13} & 0.6781 & 0.7358  \\
 		Yih et al. (2013) \cite{yih-EtAl:2013:ACL2013} & 0.7092 & 0.7700  \\ \hline
 		aNMM-2 & \textbf{0.7407} & 0.7969 \\
 		aNMM-1 & 0.7385 & \textbf{0.7995} \\\hline
 	\end{tabular}
 	\label{tab:trecqares_withno_af_fe}
 \end{table}
 
 Table \ref{tab:trecqares_withno_af_fe} shows the comparison between aNMM with previous methods using feature engineering on TRAIN-ALL without combining additional features. We find that both aNMM-1 and aNMM-2 achieve better performance comparing with other methods using feature engineering. Specifically, comparing the results of aNMM-1 with the strongest baseline by Yih et al. \cite{yih-EtAl:2013:ACL2013} based on enhanced lexical semantic models, aNMM-1 achieves $4.13$\% gain for MAP and $3.83$\% gain  for MRR. These results show that it is possible to build a uniform deep learning model such that it can achieve better performance than methods using feature engineering. To the best of our knowledge, aNMM is the first deep learning model that can achieve good performance comparing with previous methods either based on deep learning models or feature engineering for ranking answers without any additional features, syntactic parsers and external resources except for pre-trained word embeddings.
 

\subsubsection{Learning with Combining Additional Features}
Our second experimental setting is to address RQ2 proposed in Section \ref{sec:intro}, where we ask whether our model can outperform the state-of-the-art performance achieved by CNN \cite{Yu:2014, Severyn:2015:LRS:2766462.2767738} and LSTM \cite{DBLP:conf/acl/WangN15} for answer ranking when combining additional features. We combine the predicted score from aNMM-1 and aNMM-2 with the Query Likelihood (QL) \cite{Croft:2009:SEI:1516224} score using LambdaMART \cite{Wu:2010:ABI:1825381.1825396} following a similar approach to \cite{DBLP:conf/acl/WangN15}. We use the implementation of LambdaMART in jforests \footnote{\url{https://github.com/yasserg/jforests} \cite{Ganji:2011:SIGIR}.} We compare the results with previous deep learning models with additional features. Table \ref{tab:trecqares_with_af} shows the results on TRAIN and TRAIN-ALL when combining additional features. We can see that with combined features, both aNMM-1 and aNMM-2 have better performance. aNMM-1 also outperforms CNN by Severyn et al. \cite{Severyn:2015:LRS:2766462.2767738} which is the current state-of-the-art method for ranking answers in terms of both MAP and MRR on TRAIN and TRAIN-ALL.

We also tried to combine aNMM score with other additional features such as word overlap features, IDF weighted word overlap features and BM25 as in previous research  \cite{Yu:2014, Severyn:2015:LRS:2766462.2767738,DBLP:conf/acl/WangN15}. The results were either similar or worse than combining aNMM score with QL. For aNMM, the gains after combining additional features are not as large as neural network models like CNN in \cite{Severyn:2015:LRS:2766462.2767738}  and LSTM in \cite{DBLP:conf/acl/WangN15}. We think the reasons for this are two-fold: (1) The QA matching matrix in aNMM model can capture exact match information by assigning $1$ to matrix elements if the corresponding answer term and question term are the same. This exact match information includes match between numbers and proper nouns, which are highlighted in previous research work \cite{Severyn:2015:LRS:2766462.2767738}  as especially important for factoid questions answering, where most of the questions are of type \textit{what}, \textit{when} , \textit{who} that are looking for answers containing numbers or proper nouns. Within aNMM architecture, this problem has already been handled with QA matching matrix. Thus incorporating word overlap features will not help much for improving the performance of aNMM. (2) In addition to exact match information, aNMM could also learn question term importance like IDF information through question attention network. Instead of empirically designing heuristic functions like IDF, aNMM can get learning based question term importance score. As analyzed in Section \ref{sec:QueryImport}, with the optimization process in the back propagation training process, aNMM can learn similar or even better term weighting score than IDF. Thus combining aNMM score with features like IDF weighted word overlap features and BM25 may not increase the performance of aNMM by a large margin as the case in  related research works \cite{Yu:2014, Severyn:2015:LRS:2766462.2767738,DBLP:conf/acl/WangN15}. 



 \begin{table}[t]
 	\centering
 	\caption{Results of TREC QA on TRAIN and TRAIN-ALL with combining additional features.}
 	\begin{tabular}{ l || l | l  || l | l   } 
 		\hline
 		\multicolumn{1}{l||}{Training Data } & \multicolumn{2}{c||}{TRAIN} & \multicolumn{2}{c}{TRAIN-ALL}\\ \hline 
 		Method & MAP & MRR  & MAP & MRR  \\\hline
 		Yu et al. (2014) \cite{Yu:2014} & 0.7058 & 0.7800 & 0.7113 & 0.7846 \\
 		Wang et al. (2015) \cite{DBLP:conf/acl/WangN15}  & / & / & 0.7134 & 0.7913 \\
 		Severyn et al. (2015) \cite{Severyn:2015:LRS:2766462.2767738} & 0.7329 & 0.7962 & 0.7459 & 0.8078 \\\hline
 		aNMM-2 & 0.7306 & 0.7968 & 0.7484 & 0.8013 \\
 		aNMM-1 & \textbf{0.7417} & \textbf{0.8102} &\textbf{ 0.7495} & \textbf{0.8109} \\\hline
 	\end{tabular}
 	\label{tab:trecqares_with_af}
 \end{table}

\subsubsection{Results Summary}

 \begin{table}[t]
 	\centering
 	\caption{Overview of previously published systems on the QA answer ranking task. All reported results are from the best setting of each model trained on TRAIN-ALL data.}
 	\begin{tabular}{ l || l | l     } 
 		\hline
 		Method & MAP & MRR  \\\hline
	 	Wang et al. (2007) \cite{wang-smith-mitamura:2007:EMNLP-CoNLL2007} & 0.6029 & 0.6852  \\
	 	Heilman and Smith (2010) \cite{Heilman:2010:TEM:1857999.1858143} & 0.6091 & 0.6917  \\
	 	Wang and Manning (2010) \cite{Wang:2010:PTM:1873781.1873912} & 0.5951 & 0.6951  \\
	 	Yao et al. (2013) \cite{DBLP:conf/naacl/YaoDCC13} & 0.6307 & 0.7477  \\
	 	Severyn et al. (2013) \cite{DBLP:conf/emnlp/SeverynM13} & 0.6781 & 0.7358  \\
	 	Yih et al. (2013) \cite{yih-EtAl:2013:ACL2013} & 0.7092 & 0.7700  \\ \hline
	 	Yu et al. (2014) \cite{Yu:2014} & 0.7113 & 0.7846  \\
	 	Wang et al. (2015) \cite{DBLP:conf/acl/WangN15}  & 0.7134 & 0.7913  \\
	 	Severyn et al. (2015) \cite{Severyn:2015:LRS:2766462.2767738} & 0.7459 & 0.8078  \\\hline
	 	aNMM & \textbf{0.7495} & \textbf{0.8109} \\\hline
 	\end{tabular}
 	\label{tab:trecqares_all}
 \end{table}
 
 \begin{figure*}
 	\centering
 	\begin{subfigure}[b]{0.48\textwidth}
 		\centering
 		\includegraphics[width=\textwidth]{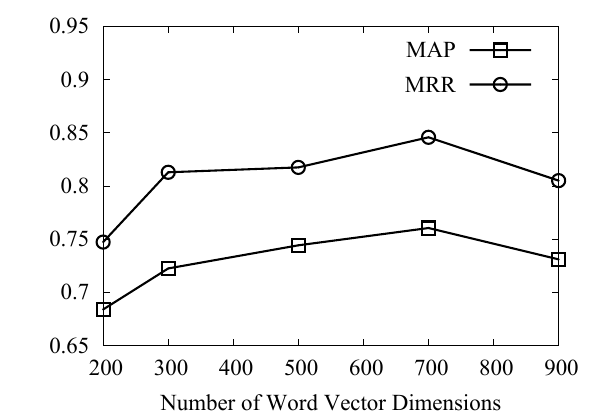}
 		\caption{Tune vector dimension on validation data.}
 		\label{fig:tune_vd}
 	\end{subfigure}
 	\hfill
 	\begin{subfigure}[b]{0.48\textwidth}
 		\centering
 		\includegraphics[width=\textwidth]{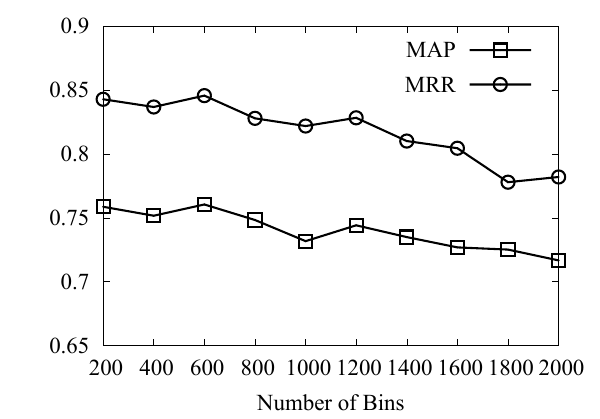}
 		\caption{Tune bin number on validation data.}
 		\label{fig:tune_bn}
 	\end{subfigure}
 	\caption{Tune hyper-parameters on validation data. }
 	\label{fig:tune_param}
 \end{figure*}

Finally we summarize the results of previously published systems on the QA answer ranking task in Table \ref{tab:trecqares_all}. We can see aNMM trained with TRAIN-ALL set beats all the previous state-of-the art systems including both methods using feature engineering and deep learning models. These results are very promising since aNMM requires no manual feature engineering, no expensive processing by various NLP parsers and no external results like large scale knowledge base except for pre-trained word embeddings. Furthermore, even without combining additional features, aNMM still performs well for answer ranking, showing significant improvements over previous deep learning model with no additional features and linguistic feature engineering methods.



%
%
%

\subsection{Parameter Sensitivity Analysis}
We perform parameter sensitivity analysis of our proposed model aNMM. We focus on aNMM-1 as the example due to the space limitation. For aNMM-1, we fix the number of bins as $600$ and change the dimension of word vectors. Similarly, we fix the dimension of word vectors as $700$ and vary the number of bins. Figure \ref{fig:tune_param} shows the change of MAP and MRR on the validation data as we vary the hyper-parameters. We summarize our observations as follows: (1) For word vector dimension, the range $(300, 700)$ is a good choice as much lower or higher word vector dimensions will hurt the performance. The choice of word vector dimension also depends on the size of training corpus. Larger corpus requires higher dimension of word vectors to embed terms in vocabulary. (2) For the number of bins, we can see that MAP and MRR will decrease as the bin number increase. Too many bins will increase the model complexity, which leads aNMM to be more likely to overfit the training data. Thus choosing suitable number of bins by optimizing hyper-parameter on validation data can help improve the performance of aNMM.

%


\section{Conclusions and Future Work}
\label{sec:conclu}

In this paper,  we propose an attention based neural matching model for ranking short answer text. We adopt value-shared weighting scheme instead of position-shared weighting scheme for combing different matching signals and incorporate question term importance learning using a question attention network. We perform a thorough experimental study with the TREC QA dataset from TREC QA tracks 8-13 and show promising results. Unlike previous methods including CNN as in \cite{Yu:2014, Severyn:2015:LRS:2766462.2767738} and LSTM as in \cite{DBLP:conf/acl/WangN15}, which only show inferior results without combining additional features, our model can achieve better performance than the state-of-art method using linguistic feature engineering without additional features. With a simple additional feature, our method can achieve the new state-of-the-art performance among current existing methods. For further work, we will study other deep learning architectures for answer ranking and extend our work to include non-factoid question answering data sets.

\section{Acknowledgments}
This work was supported in part by the Center for Intelligent Information Retrieval, in part by NSF IIS-1160894, and in part by NSF grant \#IIS-1419693. Any opinions, findings and conclusions or recommendations expressed in this material are those of the authors and do not necessarily reflect those of the sponsor.



%
%
%

\bibliographystyle{abbrv}
\bibliography{cikm}  

\begin{thebibliography}{10}

\bibitem{Croft:2009:SEI:1516224}
W.~B. Croft, D.~Metzler, and T.~Strohman.
\newblock {\em {Search Engines: Information Retrieval in Practice}}.
\newblock Addison-Wesley Publishing Company, USA, 1st edition, 2009.

\bibitem{citeulike:9606736}
O.~Etzioni.
\newblock {Search needs a shake-up}.
\newblock {\em Nature}, 476(7358):25--26, Aug. 2011.

\bibitem{Ganji:2011:SIGIR}
Y.~Ganjisaffar, R.~Caruana, and C.~Lopes.
\newblock Bagging gradient-boosted trees for high precision, low variance
  ranking models.
\newblock In {\em SIGIR '11}, pages 85--94, New York, NY, USA, 2011. ACM.

\bibitem{DBLP:conf/emnlp/GaoPGHD14}
J.~Gao, P.~Pantel, M.~Gamon, X.~He, and L.~Deng.
\newblock Modeling interestingness with deep neural networks.
\newblock In {\em Proceedings of the 2014 Conference on Empirical Methods in
  Natural Language Processing, {EMNLP} 2014, October 25-29, 2014, Doha, Qatar,
  {A} meeting of SIGDAT, a Special Interest Group of the {ACL}}, pages 2--13,
  2014.

\bibitem{Heilman:2010:TEM:1857999.1858143}
M.~Heilman and N.~A. Smith.
\newblock Tree edit models for recognizing textual entailments, paraphrases,
  and answers to questions.
\newblock In {\em Human Language Technologies: The 2010 Annual Conference of
  the North American Chapter of the Association for Computational Linguistics},
  HLT '10, pages 1011--1019, Stroudsburg, PA, USA, 2010. Association for
  Computational Linguistics.

\bibitem{NIPS2014_5550}
B.~Hu, Z.~Lu, H.~Li, and Q.~Chen.
\newblock Convolutional neural network architectures for matching natural
  language sentences.
\newblock In Z.~Ghahramani, M.~Welling, C.~Cortes, N.~D. Lawrence, and K.~Q.
  Weinberger, editors, {\em Advances in Neural Information Processing Systems
  27}, pages 2042--2050. Curran Associates, Inc., 2014.

\bibitem{Huang:2013:LDS:2505515.2505665}
P.-S. Huang, X.~He, J.~Gao, L.~Deng, A.~Acero, and L.~Heck.
\newblock Learning deep structured semantic models for web search using
  clickthrough data.
\newblock In {\em Proceedings of the 22nd ACM International Conference on
  Information \& Knowledge Management}, CIKM '13, pages 2333--2338, New York,
  NY, USA, 2013. ACM.

\bibitem{Iyyer:Boyd-Graber:Claudino:Socher:Daume-2014}
M.~Iyyer, J.~Boyd-Graber, L.~Claudino, R.~Socher, and H.~{Daum\'e III}.
\newblock A neural network for factoid question answering over paragraphs.
\newblock In {\em EMNLP '14}, 2014.

\bibitem{jansen_discourse_2014}
P.~Jansen, M.~Surdeanu, and P.~Clark.
\newblock Discourse {Complements} {Lexical} {Semantics} for {Non}-factoid
  {Answer} {Reranking}.
\newblock In {\em Proceedings of ACL'14}, pages 977--986.

\bibitem{KalchbrennerACL2014}
N.~Kalchbrenner, E.~Grefenstette, and P.~Blunsom.
\newblock A convolutional neural network for modelling sentences.
\newblock {\em Proceedings of the 52nd Annual Meeting of the Association for
  Computational Linguistics}, June 2014.

\bibitem{Keikha:2014:EAP:2600428.2609485}
M.~Keikha, J.~H. Park, and W.~B. Croft.
\newblock {Evaluating Answer Passages Using Summarization Measures}.
\newblock In {\em Proceedings of SIGIR'14}, 2014.

\bibitem{keikha_retrieving_2014}
M.~Keikha, J.~H. Park, W.~B. Croft, and M.~Sanderson.
\newblock Retrieving {Passages} and {Finding} {Answers}.
\newblock In {\em Proceedings of ADCS'14}, pages 81--84, 2014.

\bibitem{NIPS2013_5019}
Z.~Lu and H.~Li.
\newblock A deep architecture for matching short texts.
\newblock In C.~J.~C. Burges, L.~Bottou, M.~Welling, Z.~Ghahramani, and K.~Q.
  Weinberger, editors, {\em Advances in Neural Information Processing Systems
  26}, pages 1367--1375. Curran Associates, Inc., 2013.

\bibitem{NIPS2013_5021}
T.~Mikolov, I.~Sutskever, K.~Chen, G.~S. Corrado, and J.~Dean.
\newblock Distributed representations of words and phrases and their
  compositionality.
\newblock In C.~J.~C. Burges, L.~Bottou, M.~Welling, Z.~Ghahramani, and K.~Q.
  Weinberger, editors, {\em Advances in Neural Information Processing Systems
  26}, pages 3111--3119. Curran Associates, Inc., 2013.

\bibitem{DBLP:conf/aaai/PangLGXWC16}
L.~Pang, Y.~Lan, J.~Guo, J.~Xu, S.~Wan, and X.~Cheng.
\newblock Text matching as image recognition.
\newblock In {\em Proceedings of the Thirtieth {AAAI} Conference on Artificial
  Intelligence, February 12-17, 2016, Phoenix, Arizona, {USA.}}, pages
  2793--2799, 2016.

\bibitem{QiuH15}
X.~Qiu and X.~Huang.
\newblock Convolutional neural tensor network architecture for community-based
  question answering.
\newblock In {\em Proceedings of the Twenty-Fourth International Joint
  Conference on Artificial Intelligence, IJCAI 2015, Buenos Aires, Argentina,
  July 25-31, 2015}, pages 1305--1311. AAAI Press, 2015.

\bibitem{DBLP:conf/emnlp/SeverynM13}
A.~Severyn and A.~Moschitti.
\newblock Automatic feature engineering for answer selection and extraction.
\newblock In {\em EMNLP '13}, pages 458--467, 2013.

\bibitem{Severyn:2015:LRS:2766462.2767738}
A.~Severyn and A.~Moschitti.
\newblock Learning to rank short text pairs with convolutional deep neural
  networks.
\newblock In {\em Proceedings of the 38th International ACM SIGIR Conference on
  Research and Development in Information Retrieval}, SIGIR '15, pages
  373--382, New York, NY, USA, 2015. ACM.

\bibitem{DBLP:conf/cikm/ShenHGDM14}
Y.~Shen, X.~He, J.~Gao, L.~Deng, and G.~Mesnil.
\newblock A latent semantic model with convolutional-pooling structure for
  information retrieval.
\newblock In {\em Proceedings of the 23rd {ACM} International Conference on
  Conference on Information and Knowledge Management, {CIKM} 2014, Shanghai,
  China, November 3-7, 2014}, pages 101--110, 2014.

\bibitem{suGatingNN}
M.~Su and M.~Basu.
\newblock Gating improves neural network performance.
\newblock In {\em Neural Networks, 2001. Proceedings. IJCNN '01. International
  Joint Conference on}, volume~3, pages 2159--2164 vol.3, 2001.

\bibitem{Sun:2015:ODQ:2736277.2741651}
H.~Sun, H.~Ma, W.-t. Yih, C.-T. Tsai, J.~Liu, and M.-W. Chang.
\newblock Open domain question answering via semantic enrichment.
\newblock In {\em Proceedings of the 24th International Conference on World
  Wide Web}, WWW '15, pages 1045--1055, New York, NY, USA, 2015. ACM.

\bibitem{Surdeanu08learningto}
M.~Surdeanu, M.~Ciaramita, and H.~Zaragoza.
\newblock {Learning to rank answers on large online QA collections}.
\newblock In {\em ACL '08}, pages 719--727, 2008.

\bibitem{Surdeanu:2011:LRA:2000517.2000520}
M.~Surdeanu, M.~Ciaramita, and H.~Zaragoza.
\newblock Learning to rank answers to non-factoid questions from web
  collections.
\newblock {\em Comput. Linguist.}, 37(2):351--383, June 2011.

\bibitem{Tymoshenko:2015:AIS:2806416.2806490}
K.~Tymoshenko and A.~Moschitti.
\newblock Assessing the impact of syntactic and semantic structures for answer
  passages reranking.
\newblock In {\em CIKM '15}, pages 1451--1460, New York, NY, USA, 2015. ACM.

\bibitem{DBLP:conf/acl/WangN15}
D.~Wang and E.~Nyberg.
\newblock A long short-term memory model for answer sentence selection in
  question answering.
\newblock In {\em ACL '15}, pages 707--712, 2015.

\bibitem{Wang:2010:PTM:1873781.1873912}
M.~Wang and C.~D. Manning.
\newblock Probabilistic tree-edit models with structured latent variables for
  textual entailment and question answering.
\newblock In {\em Proceedings of the 23rd International Conference on
  Computational Linguistics}, COLING '10, pages 1164--1172, Stroudsburg, PA,
  USA, 2010. Association for Computational Linguistics.

\bibitem{wang-smith-mitamura:2007:EMNLP-CoNLL2007}
M.~Wang, N.~A. Smith, and T.~Mitamura.
\newblock What is the {J}eopardy model? a quasi-synchronous grammar for {QA}.
\newblock In {\em Proceedings of the EMNLP-CoNLL}, pages 22--32, Prague, Czech
  Republic, 2007. Association for Computational Linguistics.

\bibitem{Wu:2010:ABI:1825381.1825396}
Q.~Wu, C.~J. Burges, K.~M. Svore, and J.~Gao.
\newblock Adapting boosting for information retrieval measures.
\newblock {\em Inf. Retr.}, 13(3):254--270, June 2010.

\bibitem{xue_retrieval_2008}
X.~Xue, J.~Jeon, and W.~B. Croft.
\newblock Retrieval {Models} for {Question} and {Answer} {Archives}.
\newblock In {\em Proceedings of SIGIR'08}, pages 475--482, 2008.

\bibitem{DBLP:conf/ecir/YangASCPCGS16}
L.~Yang, Q.~Ai, D.~Spina, R.~Chen, L.~Pang, W.~B. Croft, J.~Guo, and
  F.~Scholer.
\newblock Beyond factoid {QA:} effective methods for non-factoid answer
  sentence retrieval.
\newblock In {\em Advances in Information Retrieval - 38th European Conference
  on {IR} Research, {ECIR} 2016, Padua, Italy, March 20-23, 2016. Proceedings},
  pages 115--128, 2016.

\bibitem{DBLP:conf/naacl/YaoDCC13}
X.~Yao, B.~V. Durme, C.~Callison{-}Burch, and P.~Clark.
\newblock Answer extraction as sequence tagging with tree edit distance.
\newblock In {\em Human Language Technologies: Conference of the North American
  Chapter of the Association of Computational Linguistics, Proceedings, June
  9-14, 2013, Westin Peachtree Plaza Hotel, Atlanta, Georgia, {USA}}, pages
  858--867, 2013.

\bibitem{yih-EtAl:2013:ACL2013}
W.-t. Yih, M.-W. Chang, C.~Meek, and A.~Pastusiak.
\newblock Question answering using enhanced lexical semantic models.
\newblock In {\em ACL '13}, pages 1744--1753, Sofia, Bulgaria, August 2013.
  Association for Computational Linguistics.

\bibitem{yin-schutze:2015:ACL-IJCNLP}
W.~Yin and H.~Sch\"{u}tze.
\newblock Multigrancnn: An architecture for general matching of text chunks on
  multiple levels of granularity.
\newblock In {\em ACL '15}, pages 63--73, Beijing, China, July 2015.
  Association for Computational Linguistics.

\bibitem{Yu:2014}
L.~Yu, K.~M. Hermann, P.~Blunsom, and S.~Pulman.
\newblock {Deep Learning for Answer Sentence Selection}.
\newblock In {\em {NIPS Deep Learning Workshop}}, Dec. 2014.

\end{thebibliography}

\end{document}